\documentclass[twocolumn,aps,prb,floatsfix,showpacs]{revtex4}
\usepackage{graphics}
\usepackage{amssymb,amsmath}
\usepackage{array}
\usepackage{dcolumn}

\begin{document}

\title{Electronic structure of the
$S\!r_{0.4}C\!a_{13.6}C\!u_{24}O_{41}$ incommensurate compound}

\author{Alain Gell\'e, Marie-Bernadette Lepetit}
\affiliation{Laboratoire de Physique Quantique, IRSAMC~/~UMR~5626,
Universit\'e Paul Sabatier, 118 route de Narbonne, F-31062 Toulouse
Cedex 4, FRANCE}

\date{\today}

\begin{abstract}
We extracted, from strongly-correlated ab-initio calculations, a
complete model for the chain subsystem of the
$S\!r_{0.4}C\!a_{13.6}C\!u_{24}O_{41}$ incommensurate compound. A
second neighbor $t-J+V$ model has been determined as a function of the
fourth crystallographic parameter $\tau$, for both low and room
temperature crystallographic structures. The analysis of the obtained
model shows the crucial importance of the structural modulations on
the electronic structure through the on-site energies and the magnetic
interactions. The structural distortions are characterized by their
long range effect on the cited parameters that hinder the reliability
of analyses such as BVS. One of the most striking results is the
existence of antiferromagnetic nearest-neighbor interactions for
metal-ligand-metal angles of $90^\circ$.  A detailed analysis of the
electron localization and spin arrangement is presented as a function
of the chain to ladder hole transfer and of the temperature. The
obtained spin arrangement is in agreement with antiferromagnetic
correlations in the chain direction at low temperature.
\end{abstract}

\maketitle

\section{Introduction}
One dimensional systems have attracted a lot of attention in the last
decade. Indeed, they present specific properties due to the constrain
imposed to electronic motion in two of the dimensions. One of their
main characteristic is the nature of their low energy excitations that
cannot be described in term of quasi particle as in the Fermi liquid
theory. On one hand, a great deal of theoretical work have been done
in order to predict the low energy physics of such systems~\cite{LL},
in particular on chain and ladder systems~\cite{ech}.  On the other
hand, solid state chemists have synthesized new compounds in order to
test the theoretical predictions.

The so-called telephone number compounds,
$S\!r_{14-x}C\!a_{x}C\!u_{24}O_{41}$, are of special interest in this
context since they present both chain and ladder
subsystems~\cite{struct1}. For instance the $x=13.6$ compound presents
under pressure a superconductive phase~\cite{ObsSupra} believed to be
a possible realization of the theoretically predicted
superconductivity in ladder systems~\cite{TheoEchSupra}. This
compound, however, is quite intriguing since it does not present the
expected properties, specific either of its chain or ladder
subsystems.  In particular, the isovalent substitution of strontium by
calcium counter-ions totally modify the magnetic low temperature
properties. Indeed, while the $x=0$ compound exhibits formation of
second-neighbors spin dimers in the chain subsystem, these dimers
disappear with increasing $C\!a$ content and an antiferromagnetic
phase appears for highly doped systems.

The $S\!r_{14-x}C\!a_{x}C\!u_{24}O_{41}$ compounds are layered systems
composed of alternated slabs of spin chains and spin ladders. The chains
and ladders run in the same direction, namely $\vec c$, however their
respective unit cell parameter in this direction are
incommensurate. The consequence of this ``misfit'' is a reciprocal
distortion of the two subsystems under the influence of electrostatic
and steric effects of the other. Indeed, each subsystems presents a
structural incommensurate modulation with the periodicity of the other
subsystem. This effect is specially important in the chain
subsystem where the $C\!u$--$O$--$C\!u$ angular distortion can reach
$10^\circ$ from the nominal value of $90^\circ$ (from $88^\circ$ to
$99^\circ$), and the copper-copper distances varies from $2.48~\rm \AA$ up
to $2.98~\rm \AA$.

Unlike what was currently assumed, we recently showed ---~from
ab-initio calculations~--- that these modulations are crucial in order
to understand the low energy physics of these
compounds~\cite{srca}. The detailed study of the undoped
compound~\cite{sr14} revealed that its specific modulations are
responsible for the electron localization observed in the chain
subsystem, the formation of second-neighbor dimers and the existence
of Zhang-Rice singlets. It has also been possible to explain the
relative occurrence of free spins and dimers observed in the magnetic
susceptibility measurements~\cite{Magn96B}.

The present paper will present an extensive ab-initio study of a
highly calcium-doped phase, namely the
$S\!r_{0.4}C\!a_{13.6}C\!u_{24}O_{41}$ compound and a model of
its chain subsystem taking explicitly into account the incommensurate
modulations. Let us remind that the
$S\!r_{14-x}C\!a_{x}C\!u_{24}O_{41}$ systems have an intrinsic doping
of 6 holes per formula unit (f.u.). The relative repartition of these
holes between the chain and ladder subsystems is an opened question,
that is believed to be of importance for the superconducting
properties. The experimental results agree on the fact that an
increasing part of the holes are located on the ladder subsystem as
the calcium content increases. However, they disagree on the actual
number of transferred holes. In the highly-doped compounds, the
evaluations of the number of holes per f.u. on the ladders go from
$1.1$ hole in the X-ray experiments~\cite{XRay00} up to $2.8$ holes in
optical conductivity experiments~\cite{COpt97}. In addition,
NMR~\cite{RMN03} experiments show an important variation of the number
of transferred holes as a function of the temperature.  We will thus
attach a special interest to this question in the present work.

The next section will be devoted to computational method, in section
III we will present and analyze a second neighbor $t-J+V$ model
Hamiltonian for the chain subsystem, extracted from the ab-initio
calculations. This model has been determined using the
crystallographic structures both at room temperature~\cite{300K} and
at low temperature ($5~\rm K$)~\cite{5K}. Section IV will discuss the
chain filling and hole transfer to the ladders as well as the magnetic
order. Finally the last section will be devoted to the conclusion.

\section{The ab-initio method} 


In strongly correlated systems, the magnetic and transfer interactions
are essentially local. They can thus be evaluated using local but
precise methods, able to treat strong correlation effects. This can be
done, for instance, using embedded-fragment ab-initio spectroscopy
methods~\cite{revue}, that allow to take advantage of the quantum
chemical multi-reference correlated codes. Indeed, we used in the
present work the Difference-Dedicated Configuration Interaction
method~\cite{DDCI} that is able to treat properly ---~within a finite
system~--- (i) the open-shell character of the magnetic/hole orbitals, (ii) the
strong electronic correlation of these orbitals, (iii) as well as the
screening effects.  The effects of the rest of the crystal on the
selected fragment are treated through an appropriate bath of charges
(Madelung potential) and total ion pseudo-potentials~\cite{TIPS}
(exclusion effect).  Such approaches have been successfully used to
study systems such as the high $T_c$ copper oxides~\cite{DDCIhtc} or
the famous $\alpha^\prime N\!aV_2O_5$ compound~\cite{vana}.

\begin{figure}[h] 
\resizebox{8cm}{!}{\includegraphics{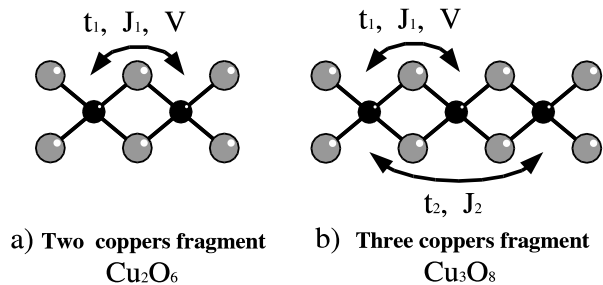}}
\caption{a) Schematic representation of the computed fragments. a) two
centers b) three centers. The gray circles represent oxygen atoms,
while the black circles represent the copper atoms.}
\label{f:frag}
\end{figure}

The quantum fragments are defined so that to include (i) the magnetic
centers, (ii) the bridging oxygens mediating the interactions, and
(iii) the first coordination shell of the preceding atoms which is
responsible for the essential part of the screening
effects~\cite{revue}. First neighbor interactions are thus determined
using $C\!u_2 O_6$ fragments (see figure~\ref{f:frag}a), while
second-neighbor ones are computed using $C\!u_3 O_8$ fragments (see
figure~\ref{f:frag}b). 
The different parameters of the $t-J+V$ model Hamiltonian are
extracted from the ab initio calculations so that to reproduce both the
computed excitation energies and the associated wave-functions,
projected onto the magnetic orbitals.
NN exchange, $J_1$, is directly given by the singlet-triplet
excitation energy when two magnetic electrons are considered in the
small fragment. NN hopping, $t_1$, and magnetic orbital energy
differences between adjacent sites, $\delta \varepsilon$, are
extracted from the first two doublet states of the same fragment with
one electron less. NNN exchange interactions, $J_2$, are extracted from
the doublet-quartet excitations energies of the three-centers
fragments with 3 magnetic electrons. NNN hopping and first neighbor
bi-electronic Coulomb repulsion are obtained from the 3 singlets and 3
triplets of same fragment with one magnetic electron less. Let us
point out that the 3 centers calculations also yield the NN
interactions. The comparison between the evaluations of the
first-neighbor integrals obtained from the 2 centers and 3 centers
fragments allows us to verify the relevance of the chosen model and
the fragment size dependence of our calculations. Some deviations can
however occur since for the three centers fragments, the calculations
where too large and we have been forced to restrain the set of treated
screening excitations by deleting the most energetic ones. In practice,  
we have excluded from the calculations all excitations going from the
deepest occupied atomic orbitals ($\varepsilon < -80 \, \rm eV$) and toward the
most energetic unoccupied orbitals ($\varepsilon > 80 \, \rm eV$).

A least-square fit method is finally used in order to extract the effective
parameters from the ab-initio calculations.

Calculations~\cite{bases} were done on 11 equivalent fragments located at 11
successive positions in the chain direction. 
These 11 fragments give a 
quite good representation of the different distortions occurring on
the chain subsystem. In order to fully represent the whole chain
subsystem, these 11 sets of results were extrapolated, using Fourier's
series analysis, as a function of the crystallographic fourth
coordinate $\tau$, associated with the system incommensurate
modulations.

Let us notice that in a complete crystallographic
description~\cite{struct4}, each chain atom possesses a fourth
fractional coordinate $\tau_i = {\vec r_i \cdot \vec k} = z_i \,
c_c/c_l$, where $\vec r_i$ is the atom position, $ c_c$ and $c_l$ the
unit cell parameters of the chain and ladder subsystem in the $\vec c$
direction, ${\vec k}= {\vec c_c^\star} \, c_c/c_l$ is the chain modulation
vector, $z_i$ is the fractional coordinate of the atom in the $\vec c$
direction. In the model Hamiltonian used in present work, $\tau$
corresponds to the fourth coordinate of the chain unit-cell copper
atom. It is defined except for a constant.

\section{A second neighbor $t-J+V$ model} 
In the chain subsystem the spins are supported by the $3d_{xz}$
orbitals of the copper atoms (the $x$ and $z$ axes being taken
respectively parallel to the $\vec a$ and $\vec c$ crystallographic
translation vectors). The model Hamiltonian is thus supported by these
copper orbitals and taken as a second neighbor $t-J+V$ model. Indeed,
the strongly correlated character of the $3d$ orbitals exclude the
possibility for the $3d_{xz}$ orbitals to be doubly-occupied and thus
suggest a $t-J+V$ model.  At the same time, the nearly $90^\circ$
$C\!u$--$O$--$C\!u$ angle is responsible for a strong hindering of
the super-exchange mechanism and thus of weak first neighbor
interactions. Second neighbor interactions are thus competitive with
the NN ones and must be taken explicitly into account.

\subsection{The Zhang-Rice singlets}
The first question we would like to address is the nature of the
orbitals supporting the holes. Indeed, it has been
suggested~\cite{XRay00} that the holes and the spins could be supported by
different orbitals and form Zhang-Rice singlets~\cite{ZR}. On the
undoped compound ($x=0$) we have been able to check this hypothesis
and show that while the spins where essentially supported by the
copper $3d_{xz}$ orbitals, the holes where essentially supported by a
copper-centered oxygen $2p$ orbital with the same local
symmetry~\cite{sr14}.

In the present compound we performed the same analysis for all the
computed two-centers fragments and the results appear to be very
similar. Indeed, while the magnetic orbitals supporting the spins have
an average $71\ \%$ copper $3d_{xz}$ content and $29\ \%$ oxygen $2p$
content, the orbitals supporting the holes have an average $12\ \%$
copper $3d_{xz}$ content and $\ 88\%$ oxygen $2p$ content. The
variations around these values are weak with standard deviations of at
most $3\ \%$.

It is noticeable that the composition of the spin/hole orbitals  is 
very stable,  as a function of \begin{itemize} \itemsep -0.5ex
\item the structural modulation, 
\item the temperature, 
\item the calcium content. 
\end{itemize}
This result is somewhat surprising, since one could have expected an
influence, on the composition of the hole orbitals, of the variations
of the $C\!u$--$O$ distances and of the rotation of the
$O$--$C\!u$--$O$ planes around the $\vec c$ axis. Indeed, on one hand,
the $C\!u$--$O$ distances present a relative variation range of $16\
\%$ in the room temperature structure and of $19\ \%$ in the low
temperature structure. On the other hand, the orientations of the
$O$--$C\!u$--$O$ planes around the $\vec c$ axis vary with a maximal
amplitude of $30^\circ$. Both parameters are expected to strongly
affect the overlap between the copper $d_{xz}$ orbital and the oxygen
$2p$ orbitals, however it seems that they do not affect the
composition of both the spin- and hole-supporting orbitals.

\subsection{The fourth coordinate fit}
The eleven sets of computed results were fitted as a function of
the fourth crystallographic coordinate $\tau$, using a Fourier series,
according to the following expression
\begin{eqnarray} \label{eq:fit}
&& a_0 + \sum_n a_n \cos{\left[2\pi n (\tau - \varphi_n) \right]} 
\end{eqnarray}
Only terms with a non negligible contribution to the series were 
retained. 

The results are summarized in tables~\ref{t:fits}a for the $5~K$
structure and~\ref{t:fits}b for the $300~K$ calculations.  Let us
notice that the parameters of the present compound necessitate more
harmonics to be fitted than for the $x=0$ compound. This can be
attributed to the larger structural modulations caused by the chemical
pressure induced by the $C\!a$ doping.

\begin{table}[ht]

(a) \begin{tabular}[t]{r|@{\hspace{1ex}}*{5}{r@{\hspace{3ex}}}r}
$5\ K$ & \multicolumn{1}{c}{$\varepsilon$ } &
\multicolumn{1}{c}{$V$} & \multicolumn{1}{c}{$t_1$} &
\multicolumn{1}{c}{$t_2$} & \multicolumn{1}{c}{$J_1$} &
\multicolumn{1}{c}{$J_2$ } \\ \hline
$a_0$  & 0 & 578 & 69.1 & 167.9 & 8.99 & -4.80 \\	   
$a_1$  & 1001 &  & -132.3 & -61.6 & -1.38 & 2.51 \\	    
$a_2$  & 574 &  & -25.9 & 24.7 & 13.70 & 0.87 \\	    
$a_3$  & &  & 31.7 & -37.5 & 0.72 & 3.07 \\		    
$a_4$  & &  & 9.8 &  & -4.41 & -0.80 \\			    
$a_5$  & &  &  &  & -2.31 &  \\                             
\\
$\varphi_1$ & 0.251 &  & 0.899 & 0.047 & 0.933 & 0.039 \\
$\varphi_2$ & 0.499 &  & 0.147 & 0.797 & 0.148 & 0.041 \\
$\varphi_3$ & &  & 0.396 & 0.213 & 0.175 & 0.551 \\
$\varphi_4$ & &  & 0.141 &  & 0.145 & 0.126 \\
$\varphi_5$ & &  &  &  & 0.998 &
\end{tabular}
\\[0.3cm]

(b) \begin{tabular}[t]{r|@{\hspace{1ex}}*{5}{r@{\hspace{3ex}}}r}
 $300\ K$& \multicolumn{1}{c}{$\varepsilon$ } &
\multicolumn{1}{c}{$V$} & \multicolumn{1}{c}{$t_1$ } &
\multicolumn{1}{c}{$t_2$} & \multicolumn{1}{c}{$J_1$ } &
\multicolumn{1}{c}{$J_2$ } \\ \hline
$a_0$  & 0 & 561 & 69.4 & 146.8 & 16.16 & -3.73 \\
$a_1$  & -1075 &  & 86.3 & 61.5 & -6.94 & 2.78 \\ 
$a_2$  & -560 &  & -72.9 & 18.8 & 1.70 & -1.18 \\ 
$a_3$  & 330 &  & 6.7 & 18.6 & -4.96 & 1.71 \\	  
$a_4$  & &  & 13.5 & 12.8 & -4.02 & 0.76 \\	  
$a_5$  & &  &  &  & -2.28 &  \\                   
\\
$\varphi_1$ & 0.761 &  & 0.406 & 0.589 & 0.382 & 0.100 \\
$\varphi_2$ & 0.246 &  & 0.648 & 0.760 & 0.684 & 0.255 \\
$\varphi_3$ & 0.584 &  & 0.406 & 0.066 & 0.399 & 0.556 \\
$\varphi_4$ & &  & 0.164 & 0.462 & 0.136 & 0.072 \\	    
$\varphi_5$ & &  &  &  & 0.413 &
\end{tabular}
\caption{Analytic fit of the $t-J+V$ second neighbor model, a) for
structure at $5\ K$, b) for structure at $300\ K$. All energies are
given in meV.}
\label{t:fits}
\end{table}

\subsection{The orbital energy}

\begin{figure}[h] 
\resizebox{8cm}{!}{\includegraphics{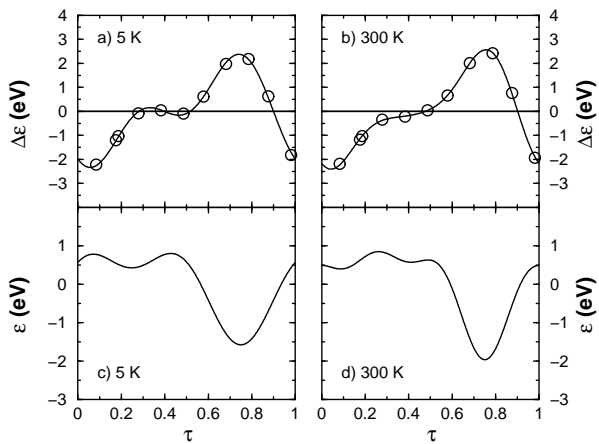}}
\caption{a),b) Energy differences ($\delta \varepsilon$) between two
NN orbitals, respectively at $5\ K$ and $300\ K$, open circles
correspond to computed values and solid line to the fit presented in
table~\ref{t:fits}. b),c) resulting orbital energies ($\varepsilon$)
respectively at $5\ K$ and $300\ K$.}
\label{f:e}
\end{figure}

The orbital energy differences between NN sites for both low and room
temperature structures are reported in figures~\ref{f:e}a and
\ref{f:e}b. The deduced orbital energies ($\varepsilon$) are reported in
figures~\ref{f:e}c and \ref{f:e}e. The variations of the orbital
energies are very large and span a range of more than $2\, \rm eV$
(respectively $2.4$ and $2.8\, \rm eV$ for the two structures). As
previously mentioned, these variations are larger than any other
interactions, and are thus responsible for the electron localization.
The electron localization can be clearly seen on the orbital energy
curves since the latter present a large potential well near
$\tau=0.75$. The widths of these well are respectively $0.45$ and $0.4$
for the $5\, K$ and $300\, K$ curves. One can thus think that about 4
electrons per f.u. will be strongly localized. Let us point out that
these 4 electrons correspond to the nominal chain occupation if all
the holes are located on them. Under the hypothesis that part of the
holes are transferred to the ladders, the corresponding electrons
transferred to the chains will be somewhat less localized due to low
on-site orbital energies. Indeed, the rest of the potential curve is
comparatively flat with an orbital energy variation of at most $0.5 \,
\rm eV$, that is only twice as large as the hopping parameters.

\begin{figure}[h] 
\resizebox{6cm}{!}{\includegraphics{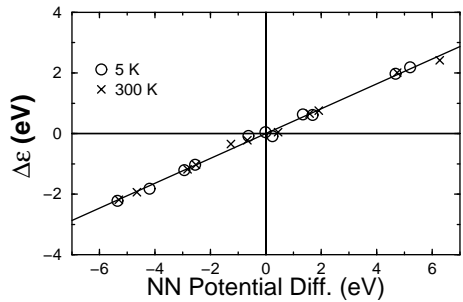}}
\caption{Orbital energy differences between neighbor copper sites as a
function of the related Madelung potential differences.  }
\label{f:mad}
\end{figure}

\begin{figure}[h] 
\resizebox{8cm}{!}{\includegraphics{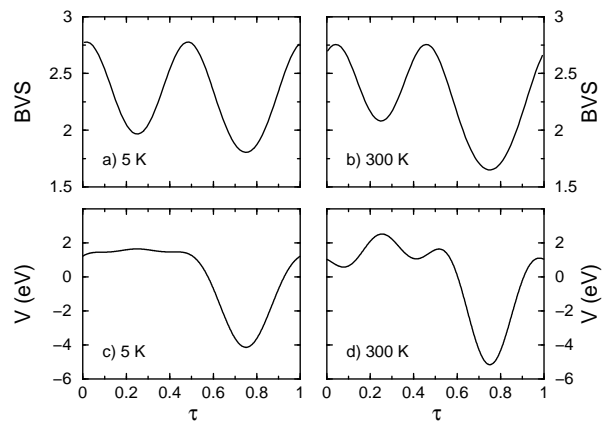}}
\caption{Copper BVS values as a function of $\tau$~: a) $T=5\ K$, b)
$T=300\ K$. Madelung potential on the copper sites~: c) $T=5\ K$, d)
$T=300\ K$. The BVS model parameters as been taken as in
reference~\cite{BVS98}, that is $r_0(C\!u-O)=1.679\rm \AA$, $B=0.37$.} 
\label{f:bvs}
\end{figure}

\begin{figure}[h] 
\resizebox{8cm}{!}{\includegraphics{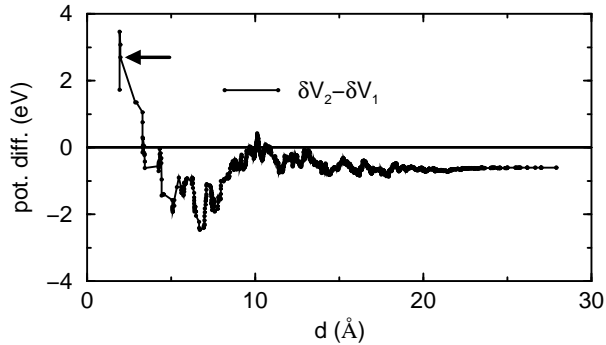}}
\caption{Contribution of the structural modulation (modulated
structure minus average structure) to the Madelung potential energy
difference between adjacent sites, labeled as 1 and 2. The
contribution is given as a function of the distances (to sites 1 and
2) up to which the modulations are taken into account. The dashed line
represent the asymptotic limit.}
\label{f:madprog}
\end{figure}

Figure~\ref{f:mad} reports the orbital energy differences between
neighboring copper atoms as a function of the related Madelung
potential differences. One see immediately, as was observed in the
undoped compound, the nearly perfect proportionality relation (with a
$0.41$ slope) between the two. This relation shows the electrostatic
origin of the orbital energy variations and the comparatively very
weak influence of the other parameters.
Let us however note that the most commonly-used tool to evaluate the
oxidation state of the atoms and thus the site occupancy is not the
Madelung potential evaluation but rather the Bond Valence Sum (BVS)
method~\cite{5K,BVS98}. This empirical technique evaluates the oxidation
state of a metal atom as a function of the distances ($r_{ij}$) of the
metal atom with its first coordination shell, that is
$${\rm BVS}_i = \sum_j \exp{((r_o - r_{ij})/B)} $$ where $r_0$ and $B$
are parameters of the model.  Figure~\ref{f:bvs} shows the Madelung
potential and the BVS evaluations for the copper sites as a function
of $\tau$. One sees immediately that while the large well present
around $\tau = 0.75$ in the Madelung potential and on-site energy
curves is correctly seen by the BVS approximation, the energetic
plateau seen for $\tau < 0.5$ is not seen in the BVS curves. In fact,
the BVS calculations exhibit another well around $\tau=0.25$. In the
low temperature curve, the BVS thus sees a quasi-doubling of the
modulation vector as in the $x=0$ compound. The use of these BVS
results in order to predict the site filling leads to the existence of
second neighbor dimeric units as in the undoped compound. Let us
remind that the experimental results do not see any dimers but an
antiferromagnetic order in this system.
The failure of the BVS method to correctly predict the electron
localization in this system can be further analyze by detailing the
contribution to the Madelung potential of a reference copper site, of
the positions modulations of the other atoms up to a given
distance.
Figure~\ref{f:madprog} reports such an analysis for two adjacent
sites (1 and 2), seen with opposite relative oxidations according
whether it is computed with the BVS method or using the Madelung
potential.
One can make the following remarks. \begin{itemize}
\item The potential difference between the two reference sites
converge slowly. Indeed, in order to reach values that fluctuate
around the correct asymptotic limit, one needs to include structural
modulations up to at least $10\, \rm \AA$.  
\item When only the first coordination shells are considered (the four
first neighbor oxygen atoms to each copper), the potential difference
between the two copper 1 and 2 has the reverse sign compared to its
asymptotic value, but the same one as the BVS. Let us remind that the
BVS only take into account this first coordination shell.  In fact,
when the modulation is applied only on the first coordination shell, the
Madelung potential and BVS methods yield similar results.
\end{itemize}
One can thus conclude that the origin the the BVS failure in the
present system is due to the long range effect of the structural
modulations on the electrostatic potential and  thus on the on-site
orbital energies.

\subsection{The first-neighbor interactions}
\begin{figure}[h] 
\resizebox{8cm}{!}{\includegraphics{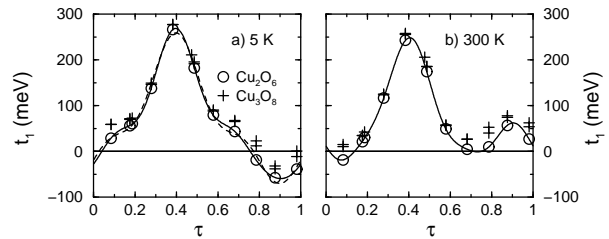}}
\caption{NN hopping integrals as a function of $\tau$. a) $5\ K$, b)
 $300\ K$.  The circles represent the computed values from the two
 copper fragments while the crosses represent the values computed from
 the three copper fragments. The solid line is the fit presented in
 table~\ref{t:fits}.}
\label{f:t1}
\end{figure}

Figure~\ref{f:t1} reports the NN hopping integrals as a function of
the fourth crystallographic coordinate $\tau$.  One sees immediately
that, both in the low and room temperature phases, the hopping
integrals present very large variations along the chain.  Indeed, it
ranges from $-60\ \rm meV$ to $270\ \rm meV$ in the low temperature
phase and over a somewhat smaller range in the room temperature
phase. Both phases present a large peak around $\tau = 0.4$, however
for $\tau \sim 0.9$ they exhibit quite different behaviors. Indeed,
in the $5\ K$ phase, the hopping integral presents a surprising
behavior since it changes sign.  

\begin{figure}[h] 
\resizebox{8cm}{!}{\includegraphics{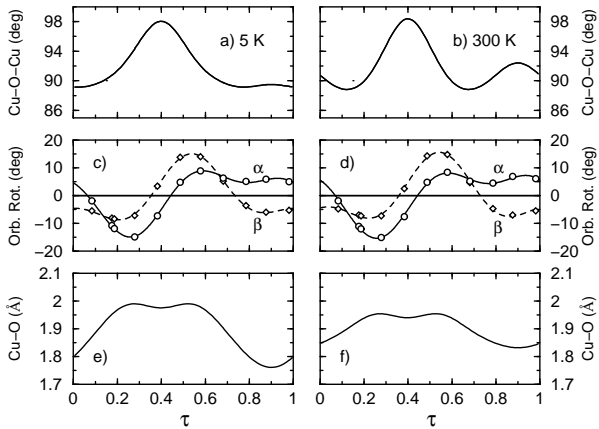}}
\caption{a) and b) $C\!u$--$O$--$C\!u$ angles . c) and d) Rotation
angle of the magnetic orbital around the $\vec c$ axis, the reference
being taken as the $C\!u$--$O_2$ plane e) and f) average
distance between  two NN $C\!u$ atoms and their two bridging-oxygen ligands.}
\label{f:dist}
\end{figure}

\begin{figure}[h] 
\resizebox{!}{8cm}{\includegraphics{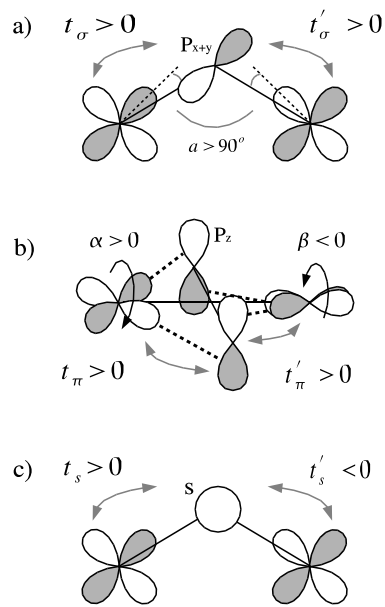}}
\caption{Schematic contribution of the different through-bridge
mechanism. a) through the oxygen $p_\sigma$ orbitals, b) through the
oxygen $2p_\pi$ orbitals, c) through the oxygen $2s$ orbitals.}
\label{f:pont}
\end{figure}

In order to explain these features, let us analyze the variation of the
hopping integral as a function of different geometric parameters (see
figure~\ref{f:dist}). The effective hopping between two magnetic
orbitals can be expressed as the sum of a direct contribution or
through-space contribution ($t_d$) and a through-bridge contribution
that goes via the bridging oxygens,
\begin{equation} \label{eq:t}
t_1 = t_d + \sum_i t_i t_i^\prime/\Delta_i
\end{equation}
where the sum runs over the different valence orbitals of the bridging
atoms and $\Delta_i$ is the ligand-to-metal charge-transfer screened
excitation energy. In the case of edge-sharing $C\!uO_2$ chains, the
through-bridge contributions are in general negligible and the only
important term is the through-space one. Indeed, the $90^\circ$
$C\!u$--$O$--$C\!u$ angles are responsible for destructive quantum
interferences as far as the in-plane oxygen $2p$ orbitals are
concerned (further referred as $2p_\sigma$). The other $2p$ orbitals
(referred as $2p_\pi$), perpendicular to the chain plane, are
orthogonal to the magnetic orbitals and thus the associated
$C\!u$--$O$ hopping terms (see figure~\ref{f:pont}b), $t_{\pi}$, are
strictly zero. Finally, the $2s$ orbitals are responsible for a
contribution which is usually very small, due to the weakness of their
overlap with the magnetic orbitals at the typical $C\!u$--$O$
distances.

Taking now into account distortions from the ideal geometry, one
generally considers that the leading perturbative term arises from the
through-bridge contribution of the in-plane oxygen $2p_\sigma$
orbitals. Indeed, this contribution is known to scale as $\cos{(a)}
\simeq a-\pi/2 $, where $a$ is the $C\!u$--$O$--$C\!u$ angle as
defined in figure~\ref{f:pont}a. 
Figure~\ref{f:dist}a and \ref{f:dist}b reports the variations of the
$C\!u$--$O$--$C\!u$ angle for the presently studied compound. One sees
immediately that the $t_1$ peak around $\tau=0.4$ is totally
correlated with the strong increase of the $C\!u$--$O$--$C\!u$ angle
at both temperatures. Figure~\ref{f:pont}a shows the mechanism
responsible for the increase of the effective hopping, $t_1$, through
the increase of the $t_\sigma^\prime$ hopping term. For angles larger
than $90^\circ$, as in the present case, $t_\sigma$ and
$t_\sigma^\prime$ have the same sign and thus contribute positively to
$t_1$.  This mechanism explains the $t_1$ peak around $\tau=0.4$,
however it cannot account for the negative peak observed in the low
temperature phases around $\tau=0.9$.

The second effect one can think of, is the set in of the $2p_\pi$
orbital contribution. Indeed, the torsion of the $C\!u$--$O_2$ local
planes compared to the reference average chain plane is responsible
for a non zero $t_\pi$ term. More precisely, if $\alpha$ is the angle
between the magnetic orbital and the $C\!u$--$O_2$ plane, $t_\pi$
scales as $\sin{(\alpha)}$.  Figures~\ref{f:dist}b and \ref{f:dist}c
report the variations of such $\alpha$ and associated $\beta$ torsion
angles (see figure~\ref{f:pont}b for the definitions) for the studied
compound. One sees immediately that unlike what was expected, there is
not any correlation between the $\alpha$ and $\beta$ angles
modulations and the low temperature peak around $\tau=0.9$. First,
both low and room temperature structure exhibit very similar
variations of $\alpha$ and $\beta$. Second, $\alpha$ and $\beta$ are
of opposite signs in the $\tau=0.9$ region and thus induce a positive
(and not the wanted negative) contribution to $t_1$.

The left over possible contribution is the $2s$ one. This
contribution, always active and of negative sign, is usually weak in
amplitude due to the too large $C\!u$--$O$ distances. Indeed, it
scales exponentially with the $C\!u$--$O$ distances. In the present
case, it can be seen in figure~\ref{f:dist}e that the $C\!u$--$O$
distances are very short around $\tau=0.9$. Let us notice that this
distortion is much weaker in the room temperature phase and its effect
is compensated by the $C\!u$--$O$--$C\!u$ angle distortion that acts
on $t_1$ with the opposite sign. In the low temperature structure the
$C\!u$--$O$--$C\!u$ angle stays close to $90^\circ$ (around $\tau = 0.9$),
thus the through-bridge contribution is essentially due to the $2s$
orbitals which act negatively on $t_1$ and are thus responsible for
the change in the hopping sign.

\begin{figure}[h] 
\resizebox{8cm}{!}{\includegraphics{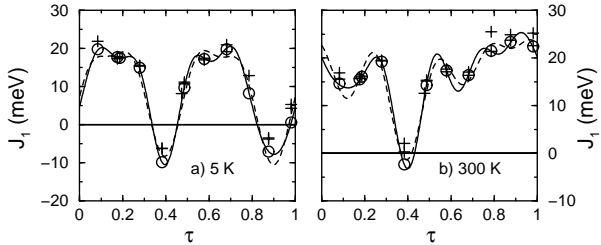}}
\caption{NN exchange integrals as a function of $\tau$. a) $5\ K$, b)
$300\ K$. The circles represent the computed values from the two
copper fragments while the crosses represent the values computed from
the three copper fragments. The solid line is the fit presented in
table~\ref{t:fits}.}
\label{f:J1}
\end{figure}

Figure~\ref{f:J1} shows the variations of the NN exchange integral at
both temperatures. One can observe that, as for the hopping, the
exchange variations are very large compared to its average
value. Indeed, the interaction even goes from ferromagnetic to
antiferromagnetic both in the low and room temperature cases.  As for
the hopping, the effective exchange $J_1$ can be expressed as the sum
of a direct or through-space contribution $J_d$ (always ferromagnetic
in nature), an Anderson super-exchange term involving the direct
hopping interactions, and through-bridge super exchange terms
associated with the three oxygen orbital types,
\begin{equation} \label{eq:J}
J_1 = J_d 
- 4 \frac{t_d^2}{U} 
- 4 \sum_i \frac{t_i^2 t_i^{\prime \, 2}}{\Delta_i^2 \, U}
\end{equation}
where $U$ is the screened one-site coulombic repulsion on the magnetic
orbital.  As expected from equation~\ref{eq:J}, a strong correlation
between the $t_1$ variations and the $J_1$ variations is observed in
figure~\ref{f:J1}. Indeed, when the geometric distortions set in, the
effective exchange integral starts to be dominated by the
through-bridge mechanisms terms and becomes antiferromagnetic. This is
the case around $\tau=0.4$ at both temperatures ---~due to the
through-bridge super-exchange mediated by the $2p_\sigma$ oxygen
orbitals~--- and for $\tau=0.9$ in the low temperature case ---~due
to the contribution of the oxygen $2s$ orbitals.

This existence of chain regions for which the NN exchange is
antiferromagnetic is, by itself, quite remarkable since it is always
assumed that NN exchange is ferromagnetic in nature in such chain
systems. In the low temperature case it corresponds to a non
negligible part of the chain, namely $27\%$. Let us
also remind that such a phenomenon has not been observed in the
undoped compound~\cite{sr14} and should thus be put into perspective
with the observation of antiferromagnetic
correlations~\cite{RMN99,Magn99,ESR01} at low temperature in the calcium highly
doped phases. This question will be discussed in somewhat more
details in the next section.

Another important remark is the existence of antiferromagnetic NN
exchange for $C\!u$--$O$--$C\!u$ angle values very close to $90^\circ$
(in the $\tau \sim 0.9$ range). This goes against all assumed models
of the magnetic interactions through $90^\circ$ angles~\cite{bac}.  Indeed, the
role of the $2s$ orbitals is usually ignored and this is the first
case to our knowledge that their effect is not only non-negligible but
dominating the interactions.

\subsection{The second-neighbor interactions}
\begin{figure}[h] 
\resizebox{8cm}{!}{\includegraphics{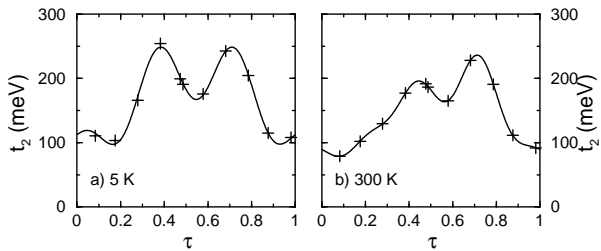}}
\caption{Second neighbor hopping integrals as a function of $\tau$. a)
$5\ K$, b) $300\ K$. The crosses represent the computed values, the
solid line represent the fit.}
\label{f:t2}
\end{figure}

Figure~\ref{f:t2} displays the effective hopping between second
neighbor magnetic orbitals. As for the $S\!r_{14}C\!u_{24}O_{41}$
undoped compound, the $t_2$ hopping integral is very large with an
average value of $167\ \rm meV$ at $5\ K$ and $147\ \rm meV$ at $300\, 
K$. The variations are large, but unlike the NN hopping $t_2$ does not
change sign. Indeed, in this case the dominating process is clearly
the through-bridge process mediated by the $2p_\sigma$ orbitals of the
four bridging oxygen atoms. 
\begin{figure}[h] 
\resizebox{8cm}{!}{\includegraphics{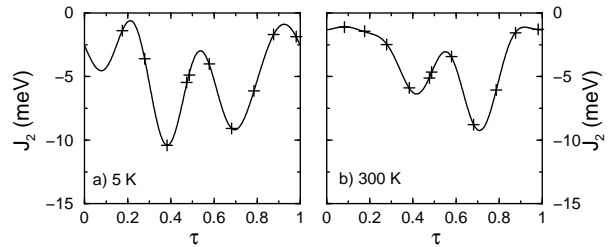}}
\caption{Second neighbor exchange integrals as a function of $\tau$. a)
$5\ K$, b) $300\ K$. The crosses represent the computed values, the
solid line represent the fit.}
\label{f:J2}
\end{figure}
Figure~\ref{f:J2} reports the second neighbor exchange integrals
between the magnetic orbitals. As expected they are antiferromagnetic,
the super-exchange mechanism being mediated by the same $2p_\sigma$
orbitals as for the hopping. Consequently, the maximal amplitude of
$J_2$ is obtained for the largest values of $t_2$, as can be seen
on figure~\ref{f:J2}.

\section{Filling analysis}
\begin{figure*}[ht] 
\resizebox{15cm}{!}{\includegraphics{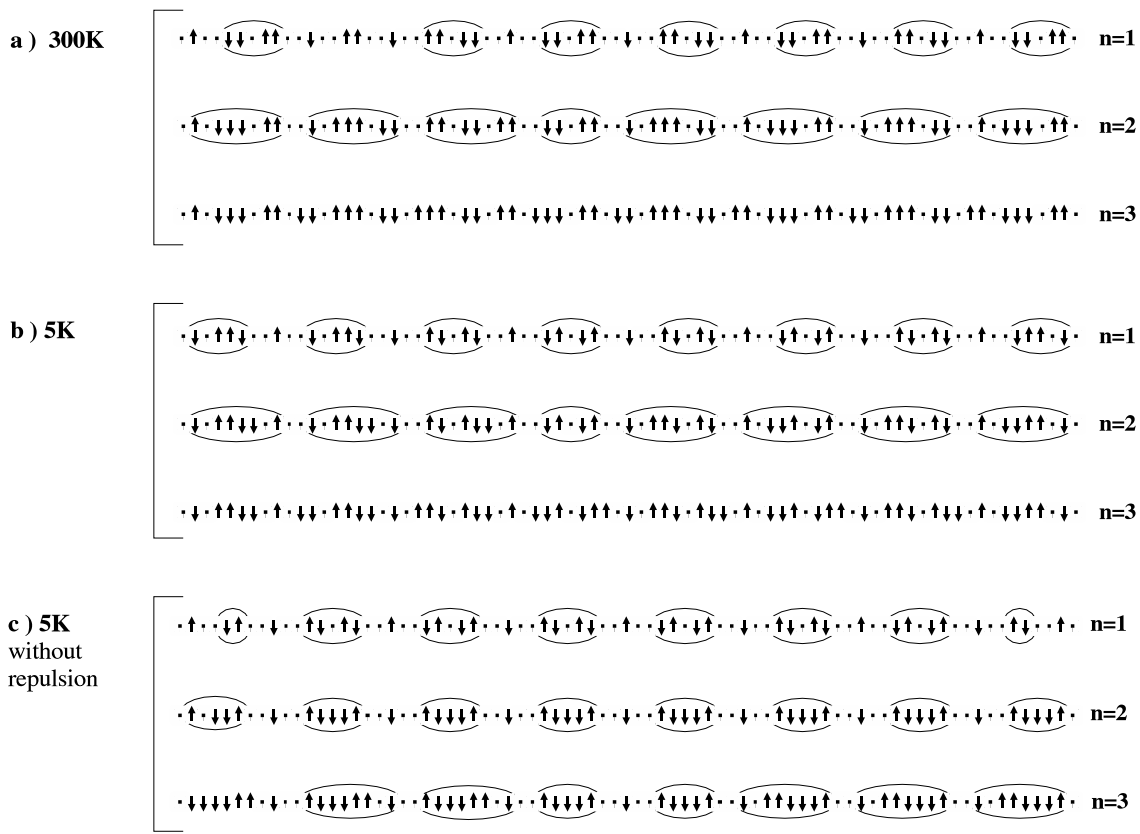}}
\caption{Tentative electron localization as a function of the filling
($n=1,\,2,\,3$ where $n$ is the number of holes per f.u. transferred
to the ladders). a) and b) at $5\ K$, c) at $300\ K$. In the a)
diagrams the localization is computed using the magnetic orbitals
on-site energies only, in the b) and c) diagrams the electron
localization is computed taking into account the NN coulombic
repulsion in addition to the on-site energies.}
\label{f:remp}
\end{figure*}

Let us now consider the possible fillings and associated electron
localization in the chain subsystem. Figure~\ref{f:remp}a displays the
localization of the electrons over 77 consecutive sites along the
chain, in the $5\ K$ structure. The electrons are localized according
to an energetic criterion taking into account both the on-site orbital
energies and the NN bi-electronic repulsion. In order to see more
clearly the effect of the bi-electronic repulsion we have reproduced
the filling obtained without the NN repulsion as published in
reference~\cite{srca} (figure).  Spins are arranged according to NN
and NNN exchange interactions. Three types of filling are considered
in the range proposed in the literature, namely with one, two and
three holes per f.u. transferred to the ladders.

One sees immediately that there is not any dimers for either of the
fillings. The second important remark, is the presence at $5\, K$ and
for all fillings, of NN occupied sites with antiferromagnetic (AF)
exchange. Indeed, the AF interactions allows the formation of clusters
with no spin frustration.  The $27\%$ of $J_1$ antiferromagnetic
values are thus directly related with the possibility to transfer
electrons to the chain subsystem without increasing the magnetic
energy.  
As can be seen from the comparison between figures~\ref{f:remp}b and
\ref{f:remp}c. the effects of the NN repulsions are (i) to increase
the typical size of the spin clusters, (ii) to prevent the appearance
of spin frustration in the clusters, while some can be observed when
repulsions are not taken into account, (iii) to favor the
disappearance of the free spins. Indeed, the latter disappear for
$n\le 2$, while $n\le 3$ is necessary if the repulsion is not
considered.  Let us notice that free spins are not experimentally
observed in this highly doped compound~\cite{Magn99}, unlike the
undoped one~\cite{Magn96B}.  

In the room temperature phase, the formation of high spin cluster,
partially frustrated is observed. Indeed, the small range of $\tau$
where the NN exchange integral is antiferromagnetic is associated with
consecutive sites supporting at least one hole. The consequence is
that unlike what happens at $5\ K$, the AF exchange parameter region
is in this case non relevant for the physics.

\section{Conclusion}
The present paper proposes a second neighbor $t-J+V$ model for the
$S\!r_{0.4}C\!a_{13.6}C\!u_{24}O_{41}$ compound, as a function of the
fourth crystallographic coordinate $\tau$. The model Hamiltonian has
been extracted from a series of ab-initio calculations and is given
both for the $5\ K$ and $300\ K$ structures. 

The present study shows the crucial importance of
the structural modulations on the low energy properties of the
compound. 
Indeed, the on-site energy of the magnetic orbitals is
strongly modulated by the atomic displacements and is affected by
geometric variations occurring at distances up to $\sim 16\ \rm \AA $. 
The consequence of such long range dependence of the on-site
energies is the non pertinence of methods such as BVS for the
evaluation of the copper oxidation state and thus of the electron
localization.
The other important effect of the structural modulations is the
appearance of large ranges of $\tau$ for which the NN exchange become
antiferromagnetic despite the nearly $90^\circ$ $C\!u$--$O$--$C\!u$
angles. In particular, it has been observed for the first time
dominant through-bridge exchange mechanism going via the oxygen $2s$
orbital.

The consequence of these particularities on the low energy properties
of this compound is a strong electron localization for a large part of
the sites, and the disappearance of spin frustration in the low
temperature phase. Another consequence is the incredible sensibility
of the electronic structure and more specifically of the localization
of the electrons to the specific distortions of the geometric structure. 

One of the main remaining question in this family of systems and in
particular in the calcium doped phases is the position of the Fermi
level or in other words the relative chemical potential of the chains
and ladders subsystems that drive the hole transfer between the
latter.  In order to answer such a question it is necessary to study
the ladder-chain interactions, the relative positions of the on-site
energies of both subsystems, the hopping between the two as well as
the chain-ladder magnetic interactions that could be responsible for a
three dimensional spin ordering. In particular, it would be 
interesting to know whether there exist large hopping terms between
ladder and chain copper sites with on-site energies both near the
Fermi level. The underlying question is to which extend these systems
should be considered as quasi one-dimensional.

{\bf Acknowledgment :} the author would like to thank Dr. J. Etrillard
for discussions on the structural aspects, Dr. D. Maynau for providing
us with the CASDI chain of programs, Dr. D. Poilblanc for helpful
discussions. The calculations were performed at IDRIS/CNRS
computational facilities under project number 1104.

 \end{document}